\documentclass{elsart3}
\usepackage{graphicx}
\begin{document}

\begin{frontmatter}
\title{Evolution of the heavy fermion state in Ce$_2$IrIn$_8$ }
\author[LANL,JAERI]{R.H.~Heffner\corauthref{cor}}
\corauth[cor]{Tel. +81-29-284-3524,
Fax: +81-29-282-5927,
email: heffner@popsvr.tokai.jaeri.go.jp} 
\author[LANL,TRI]{G.D.~Morris}
\author[LANL]{E.D. Bauer}
\author[LANL]{J.L. Sarrao}
\author[LANL]{J.D. Thompson}
\author[UCR]{ D.E.~MacLaughlin}
\author[UCR]{L.~Shu}   
\address[LANL]{Los Alamos National Laboratory, MS K764, Los Alamos, NM 87545,
             USA}
\address[UCR]{Department of Physics, University of California,
              Riverside, CA 92521, USA}
\address[JAERI]{Japan Atomic Energy Research Institute, Tokai-Mura, Naka-Gun, Ibaraki-Ken, 319-1195 Japan}
\address[TRI]{TRIUMF, 4004 Wesbrook Mall, Vancouver, B.C., Canada V6T 2A3}

\begin{abstract}
We report $\mu$SR Knight shift and susceptibility studies for $1$~T applied field along the crystalline c- and a-axes of  the heavy fermion compound Ce$_2$IrIn$_8$. Below a characteristic temperature $T^*$ one observes a `Knight-shift anomaly' in which  the Knight shift constant $K$ no longer scales linearly with $\chi$. This anomaly is consistent with a scaling law in which the  susceptibility $\chi$ is composed of a high-temperature component corresponding to non-interacting local moments and a low-temperature component $\chi_{\rm cf} \sim (1-T/T^*)\ln(T^*/T)$ which characterizes the heavy-electron state below $T^*$. We find that  $T^*$ is anisotropic, with $T_a^* = 59(3)$~K and $T_c^* = 24(1)$~K, and derive the magnitudes of $\chi_{\rm cf}^{a,c}$.
\end{abstract}

\end{frontmatter}

\section{Introduction}
It is generally agreed that a defining characteristic of f-electron heavy-fermion materials is the transformation of independently fluctuating moments at high temperatures into a system of interacting, highly screened moments at low temperatures via hybridization between the localized f-electrons and the conduction electrons. This process is understood exactly in the case of a single Kondo impurity, but there is no complete theory of the formation of the heavy fermion state in a lattice of f-electrons. Recently, a phenomenological picture for the thermodynamics of this process in the Ce$_{1-x}$La$_x$CoIn$_5$ system was put forth in terms of a two-component susceptibility \cite{Nakatsuji}: a  high-temperature component, corresponding to non-interacting, local Ce moments, and a low-temperature component $\chi_{\rm cf} \sim (1-T/T^*)\ln(T^*/T)$ which characterizes the formation of the heavy-electron state below temperature $T^*$.    In order to study the generality of this phenomenology we measured transverse field (TF) $\mu$SR Knight shifts (which measure the local susceptibility) in the heavy fermion material Ce$_2$IrIn$_8$. This material is the two-layer analog to CeIrIn$_5$ and  possesses a relatively high Sommerfeld constant of $700$ mJ/Ce mol-K$^2$ \cite{115218}. In a previous zero-field $\mu$SR investigation \cite{morris} we showed that Ce$_2$IrIn$_8$ possesses small-moment, disordered magnetic ordering below about $T = 0.6$ K. Our TF measurements reported here were carried out well above this ordering temperature, between $4 - 300$ K.

The muon Knight shift constant $K=(\nu-\nu_0)/\nu_0$, where  $\nu$ is the measured frequency  and $2\pi\nu_0 = \gamma_{\mu}H_0$, where $\gamma_{\mu}$ is the muon's gyromagnetic ratio ($8.51 \times 10^8$~Hz/T) and $H_0$ is the applied field.  Generally,  $K = K_0 + K_{\rm dem} + B_{\rm hyp}\chi_f(T)/N_A\mu_B$, where $K_{\rm dem}$ is the shift caused by the demagnetization fields, $\chi_f$ is the total temperature-dependent f-electron susceptibility, $K_0$ is the shift from temperature-independent sources and $B_{\rm hyp}$ is the total hyperfine coupling field between the muon and the Ce moments. The constants $N_A$ and $\mu_B$ are Avogadro's number and the Bohr magneton, respectively. For muons one usually has $B_{\rm hyp} = B_c + B_{\rm dip}$, where $B_c$ is an indirect RKKY-type coupling and $B_{\rm dip}$ is a direct dipolar coupling \cite{Schenck}. One expects $K$ proportional to the bulk susceptibility if $K \propto B_{\rm hyp}(r)\chi(T)$.  A breakdown of linearity between $K$ and $\chi$ is found in many heavy fermion materials \cite{cox,curro}, however, implying $K \propto B_{\rm hyp}(r,T)\chi(T)$ (i.e., a temperature-dependent  $B_{\rm hyp}$), or, as considered here, an additional low-temperature susceptibility component which couples differently to the muon.

\section{Experimental results}
\begin{figure}
\centering
\includegraphics[width=\columnwidth]{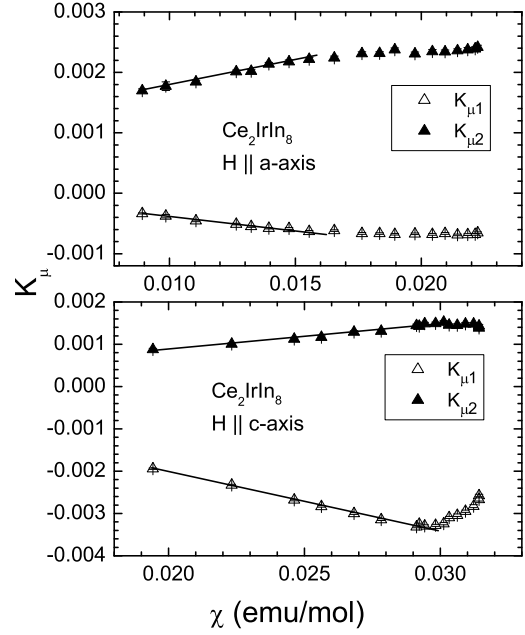}
\caption{
Susceptibility dependence of $K_{\mu} = K-K_{\rm dem}$ for applied fields along the $c$- and $a$-axes. The two frequencies correspond to two muon sites.}
\label{KvsChi}
\end{figure}

The experiments were performed at the M20 muon channel at TRIUMF in Vancouver, Canada.  Single crystal samples were mounted on a silver backing which served as a reference material to calibrate the applied field $H_0=1$~T. Fig. \ref{KvsChi} shows the susceptibility dependence of the measured $K_{\mu i}^{a,c} \equiv K_i^{a,c} - K_{\rm dem}^{a,c}$ for $H_0$ parallel to the $a$- and $c$-axes, where $K_{\rm dem} = 4\pi(\frac{1}{3}-N)\rho_{\rm mol}\chi$, with $\rho_{\rm mol}$ the molar density. The geometrical demagnetization terms  are $N\cong0.73$ and $N\cong0.82$ for the $c$- and $a$-axis data, respectively. Two frequencies were observed with shifts labeled $K_{\mu 1}$ and $K_{\mu 2}$, corresponding to muons stopping at different sites with relative amplitudes $2:3$, respectively. A non-linear $K-\chi$ relation is observed for both sites and both field directions.
\begin{figure}
\centering
\includegraphics[width=\columnwidth]{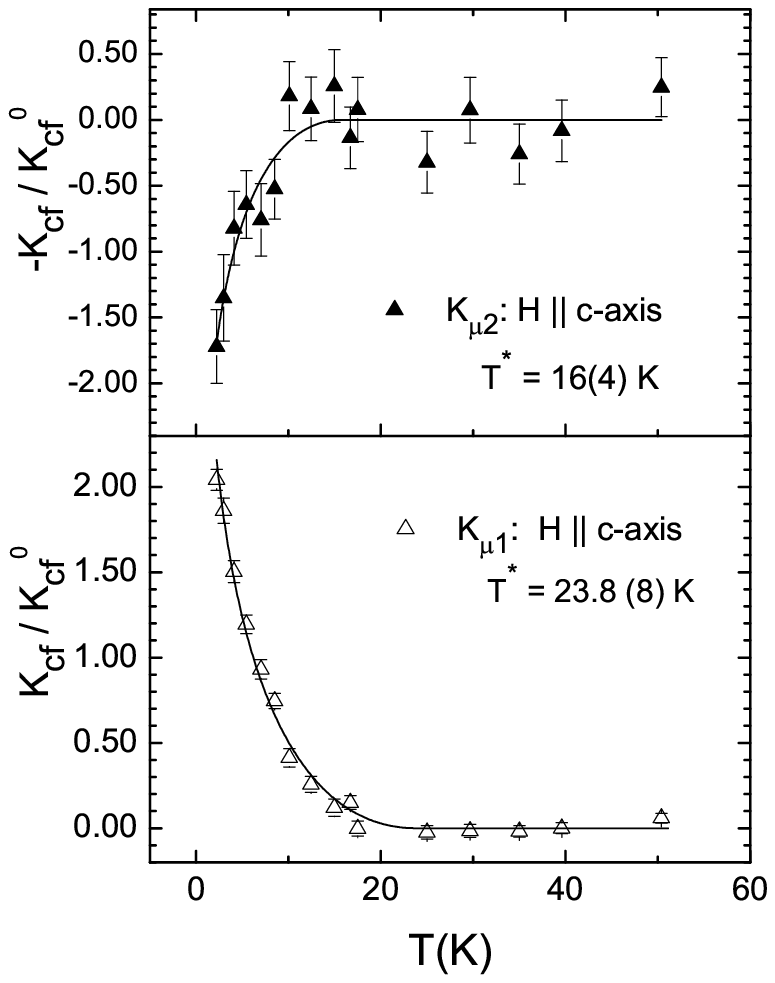}
\caption{
Temperature dependence of the normalized (see text) Knight $K_{\rm cf}$ for $H_0 \parallel c$-axis: The two shifts $K_{\mu 1}$ and $K_{\mu 2}$ correspond to two muon sites. The solid lines are fits to the equation $K_{\rm cf} = K_{\rm cf}^0 (1-T/T^*)\ln(T^*/T)$ \cite{Nakatsuji,curro}.}
\label{KCfit}
\end{figure}
\begin{figure}
\centering
\includegraphics[width=\columnwidth]{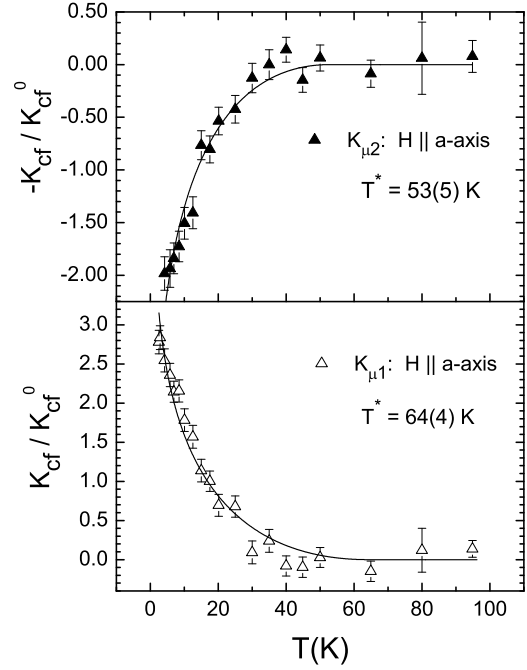}
\caption{
Temperature dependence of the normalized (see text) Knight $K_{\rm cf}$ for $H_0 \parallel a$-axis: The two shifts $K_{\mu 1}$ and $K_{\mu 2}$ correspond to two muon sites. The solid lines are fits to the equation $K_{\rm cf} = K_{\rm cf}^0 (1-T/T^*)\ln(T^*/T)$ \cite{Nakatsuji,curro}.}
\label{KAfit}
\end{figure}
Following the notation of Curro {\em et al.} \cite{curro} we write the molar susceptibility as
$\chi = \chi_{\rm ff} + 2\chi_{\rm cf} + \chi_0$, where $\chi_{\rm ff}$   corresponds to  high-temperature localized spins, and $\chi_0$ is independent of temperature. One has \cite{curro} 
\begin{eqnarray}
K_{\mu i}^{a,c} &=& [(A_i^{a,c}-B_i^{a,c})\chi_{\rm cf}^{a,c} \nonumber \\
                 & &+B_i^{a,c}(\chi^{a,c} - \chi_0^{a,c})]/N_A\mu_B, 
\label{Kchi}
\end{eqnarray}

 where $i=1,2$ and $A$ is the  coupling between the muon and $\chi_{\rm cf}$. Eq. \ref{Kchi} is actually 4 equations, one for each site and field direction. Because $\chi_{\rm cf}$ vanishes above $T^*$, the coupling constants $B_i^{a,c}$ can be determined by fitting the $K_{\mu i}^{a,c}$ vs. $\chi^{a,c}$ data at high temperatures  \cite{curro}.  This is shown by the solid lines in Fig. \ref{KvsChi}. If the muon sites possess tetragonal symmetry, then $B= B_c + B_{\rm dip}^{zz}$ for $H \parallel c$ and $B=B_c - \frac{1}{2}B_{\rm dip}^{zz}$ for $H \parallel a$. 
Here $B_{\rm dip}^{zz}$ is the $z$-component of the dipole coupling tensor and $B_c$ is assumed to be isotropic. Thus, $B_{\rm dip}^{zz}$ and $B_c$ can be determined for each site.
\begin{table}
\begin{center}
\caption{Hyperfine fields $B$  derived from fitting high-temperature $K$ vs. $\chi$ data in Fig. \ref{KvsChi} using Eq. \ref{Kchi}. Also shown is $B_{\rm dip}^{zz}$, derived assuming tetragonal symmetry for the muon sites corresponding to $K_{\mu 1}$ and $K_{\mu 2}$ (see text).}
\begin{tabular}{|c|c|c|} \hline
Hyperfine fields (Oe/$\mu_B$)   &    $K_{\mu1}$	      &    $K_{\mu2}$\\ \hline
$B(H_0 \parallel c)$  &    $-786 \pm 9$	&    $379 \pm 20$ \\ \hline
$B(H_0 \parallel a)$  &    $-248 \pm 9$	&    $475 \pm 41$ \\ \hline
$B_{\rm dip}^{zz}$    &    $-359 \pm 12$     &    $-64 \pm 40$ \\ \hline
\end{tabular}

\end{center}
\end{table}

 The $B_i^{a,c}$ from high-temperature fits to the data  are given in Table I, together with the derived values of $B_{\rm dip}^{zz}$, assuming tetragonal symmetry. Three possible tetragonal sites are $(\frac{1}{2},\frac{1}{2},\frac{1}{2})$, $(\frac{1}{2},0,0.194)$ and  $(\frac{1}{2},\frac{1}{2},0)$ (see Fig. 1 of Ref. \cite{macaluso}). The calculated $B_{\rm dip}^{zz}$ for these sites are $423, -1523$ and $-156$ Oe/$\mu_B$ , respectively. The $K_{\mu2}$ site is thus roughly consistent with the $(\frac{1}{2},\frac{1}{2},0)$ site, and we derive a value of $B_c = 443(60)$ Oe/$\mu_B$ for this site.  The $K_{\mu1}$ site, however, does not agree with any of these obvious sites and is not yet identified.

Having obtained the values of $B_i^{a,c}$ (Table I), one may proceed to solve Eq. \ref{Kchi} for the terms containing $\chi_{\rm cf}^{a,c}(T)$.  These data, shown in Figs. \ref{KCfit} and \ref{KAfit}, are well-described by $K_{\rm cf} = K_{\rm cf}^0 (1-T/T^*)\ln(T^*/T)$ \cite{curro}, where $K_{\rm cf} = K_{\rm cf}^0$ for $\alpha=T/T^*=0.259, (1-\alpha)\ln\alpha=1$.  The weighted averages over the two muon sites yield  $T_a^* = 59(3)$~K and $T_c^* = 24(1)$~K.

\begin{figure}
\centering
\includegraphics[width=\columnwidth]{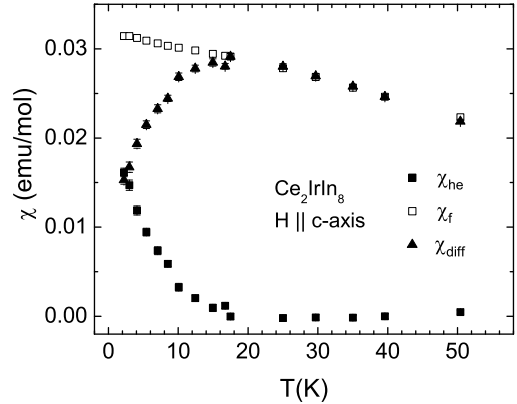}
\caption{
Temperature dependence of the susceptibilities $\chi_{\rm he}$, $\chi_{\rm f}$ and $\chi_{\rm diff}=\chi_{\rm f}-\chi_{\rm he}$ for applied fields along the $c$-axis. }
\label{ChisC}
\end{figure}

\begin{figure}
\centering
\includegraphics[width=\columnwidth]{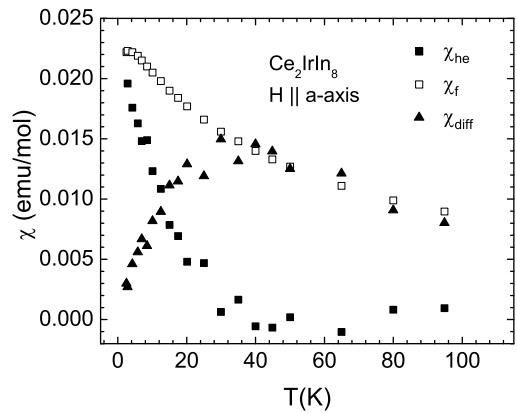}
\caption{
Temperature dependence of the susceptibilities $\chi_{\rm he}$, $\chi_{\rm f}$ and $\chi_{\rm diff}=\chi_{\rm f}-\chi_{\rm he}$ for applied fields along the $a$-axis. }
\label{ChisA}
\end{figure}

We now make  the reasonable assumption that the contact coupling $A$ is isotropic, e.g., $A_i^a = A_i^c$ for $i=1,2$. Then for $T<T^*$ the four equations corresponding to Eq. \ref{Kchi} can be solved for the $A$ values at each site and the $\chi_{\rm cf}^{a,c}$ for the $a$ and $c$ directions. This extends the analysis beyond that in Ref. \cite{curro}, and allows us to extract the value of $\chi_{\rm cf}$ in emu/mol for each field direction.  The results are shown in Figs. \ref{ChisC} and \ref{ChisA}, where we have plotted the total f-electron susceptibility $\chi_{\rm f}=\chi-\chi_0$, the heavy electron susceptibility $\chi_{\rm he} \equiv 2\chi_{\rm cf}$ \cite{curro} and $\chi_{\rm diff} \equiv \chi_{\rm f}-\chi_{\rm he}$ as a function of temperature.
The most reliable values of the hyperfine couplings $A$ were determined from the data points  below $10$~K where the shifts for $H\parallel c$  are large and statistically well determined (Figs. \ref{KCfit} and \ref{KAfit}). The $A$ values are relatively independent of temperature: $A_1=-118(28)$ Oe/$\mu_B$ and $A_2=272(22)$ Oe/$\mu_B$.
\section{Discussion}
Through a combination of $\mu$SR Knight shift and bulk susceptibility measurements we have obtained  the magnitudes of the hyperfine coupling parameters and local heavy-fermion spin susceptibilities \cite{Nakatsuji,curro} in Ce$_2$IrIn$_8$. The energy scale $T^*$ governing the evolution of the coherent heavy fermion state is anisotropic, with $T_a^* = 59$~K and $T_c^* = 24$~K. These values can be compared with the lattice Kondo energy scale $T_0$  of $31$~K derived from the Sommerfeld constant for $J=5/2$ (Ce$^{3+}$) \cite{rajan}, and the resistivity, which shows a broad maximum near $40-50$~K. Our most important result is the measurement of the magnitude and temperature dependence of $\chi_{\rm he}^{a,c}(T)$, showing the relative fraction of coherent and incoherent spectral density in the Ce spins. The extrapolated $\chi_{\rm he}^a$  approaches the total f-electron susceptibility $\chi_f^a$ for $T \approx 2$~K, but $\chi_{\rm he}^c$ does not approach $\chi_f^c$ until  $T \approx 0.5$~K. Thus, there are likely uncompensated Kondo spins for $T \leq 0.6$~K, which may be the source of the disordered spin freezing found in  our previous $\mu$SR measurements \cite{morris}.


Work at LANL performed under
 the U.S. D.O.E. Work
at Riverside  supported by the U.S. NSF, Grant DMR-0102293.
We thank the staff of TRIUMF, as well as
 N. J. Curro, Z. Fisk and D. Pines for helpful discussions.


\begin{thebibliography}{xx}
\bibitem{Nakatsuji} Satoru Nakatsuji, David Pines and Zachary Fisk, Phys. Rev. Lett. 92 (2004) 106401.
\bibitem{115218} N.O.~Moreno, M.F.~Hundley, P.G.~Pagliuso, R.~Movshovich, M.~Nicklas,
J.D.~Thompson, J.L.~Sarrao, Z.~Fisk, Physica B312--313, 274 (2002). 
\bibitem{morris}G.D.~Morris, R.H.~Heffner, N.O. Moreno, P.G. Pagliuso, J.L. Sarrao, S.R. Dunsiger, G.J. Nieuwenhuys, D.E. 
MacLaughlin and O.O. Bernal, Phys. Rev B69, 214415 (2004).
\bibitem{cox} D.E. MacLaughlin, J. Magn. Magn. Mater. 47-48, 121 (1985); E. Kim and D. L. Cox, Phys. Rev. B58, 3313 (1998).
\bibitem{curro} N. J. Curro, B.-L. Young, J. Schmalian and D. Pines, Phys. Rev B70, 235117 (2004).
\bibitem{Schenck} A.~Schenck, Muon Spin Rotation Spectroscopy: Principles and Applications in Solid State Physics (Hilger, Bristol, Boston, 1985).
\bibitem{macaluso} R.T. Macaluso, J.L.Sarrao, N.O. Moreno, P.G. Pagliuso, J.D. Thompson, F.R. Fronczek, M.F. Hundley, A. Malinowski and J.Y. Chan, Chem. Mater. 15, 1394 (2003). 
\bibitem{rajan} V.T. Rajan, Phys. Rev. Lett. 51, 308 (1983).
\end{thebibliography}
\end{document}